\documentstyle[twoside,fleqn,espcrc2,epsf]{article}

\newcommand{\AmS}{{\protect\the\textfont2
  A\kern-.1667em\lower.5ex\hbox{M}\kern-.125emS}}

\begin{document}

\title{The Phase Diagram of the $U(2)\times U(2)$ Sigma Model
\thanks{Presented by V. Koulovassilopoulos}} 

\author{D.~Espriu, V.~Koulovassilopoulos, and A.~Travesset
\address{Departament d'Estructura i Constituents de la
        Mat\`eria,  Universitat de Barcelona,\\
             and 
        Institut de F\'\i sica d'Altes Energies, 
        Diagonal, 647,  E-08028 Barcelona, Spain.}}

\begin{abstract}
We study the phase diagram of the $U(2) \times U(2)$ scalar model in
$d=4$ dimensions. We find that the phase transition is of first order in
most of the parameter space. The theory can still be relevant to 
continuum physics (as an effective theory)  provided the transition is
sufficiently weakly first order. This places restrictions on the
allowed coupling constants.
\end{abstract}

\maketitle

\section{The $U(N) \times U(N)$ model}\label{int}
We consider a scalar field theory described by the action
\begin{eqnarray}
S(\phi)&=&\int d^4x (\frac{1}{2}{\rm Tr} (\partial_{\mu}\phi^{\dagger}
\partial_{\mu}\phi)+
\frac{1}{2}m^2{\rm Tr} ( \phi^{\dagger} \phi)
 \nonumber
 \\ &&
+\lambda_1 ({\rm Tr} \phi^{\dagger} \phi)^2+
\lambda_2 {\rm Tr} ( \phi^{\dagger} \phi)^2 ),
\label{action}
\end{eqnarray}
in Euclidean space, where $\phi(x)$ is a complex $N \times N$ matrix.
The action (\ref{action})  is invariant under a $U(N)_L\times U(N)_R$
(to be taken $N=2$) symmetry, under the global symmetry transformation
$\phi \rightarrow L \phi R^\dagger$,  where $L,R$ are $U(N)$
matrices. This model has been considered as a low-energy effective
theory to describe the strong-coupling extended technicolor models and
top-condensate models of electroweak symmetry breaking \cite{CGS}.

This model, in contrast to the $O(N)$ model, is known to possess for
$\lambda_2 \neq 0$ a first order phase transition whose strength varies
in the $(\lambda_1, \lambda_2)$ space. This is due to the well-known
Coleman-Weinberg instability \cite{CW}, and it places restrictions on
the allowed parameter space of couplings.  As one adjusts $m^2$ past a
critical value, $m_c^2$, the vacuum expectation value (v.e.v.) $v$ jumps
discontinuously from zero in the unbroken phase to some finite nonzero
value in the broken phase.  
Then, if the model is to be a valid low-energy effective theory,
relevant to continuum physics, the couplings $(\lambda_1,\lambda_2)$,
should belong to a region where the phase transition is sufficiently
weak first order. It is only then that $v$ can be small compared to
the cut-off $\Lambda$.

The model (\ref{action}) has been studied in perturbation theory, in
terms of the effective potential and the renormalization group (RG) in
ref.~\cite{PATERSON} and in the language of RG flows and its associated
fixed points \cite{YAM,AMIT} in \cite{CGS,BHJ}. A preliminary
investigation on the lattice was  undertaken in ref.~\cite{YUE}.
We present here a more complete investigation of the phase structure of
the $U(2)\times U(2)$ model, by performing Monte Carlo simulations of
the lattice regularized version of the action (\ref{action})
above.

\section{Perturbation theory}

In the sequel we take $N=2$,  so the action (\ref{action}) depends on
eight degrees of freedom. More details about the model can be found
in \cite{YUE,EKT}. If $\lambda_2=0$, then the symmetry is enhanced to
$O(8)$ ($O(2N^2)$ for general $N$).
The pattern of symmetry breaking depends on the sign of $\lambda_2$. If
$\lambda_2>0$ then  the breaking occurs according to 
\begin{equation}
U(2)_L \times U(2)_R \rightarrow U(2)_V
\end{equation}
resulting in four Goldstone bosons, while if $\lambda_2 <0$, the
symmetry breaking pattern is that of
\begin{equation}
U(2)_L \times U(2)_R \rightarrow U(1)^3 \label{SB2}
\end{equation}
resulting in five Goldstone bosons.


\begin{figure}[tb]
\epsfxsize=80mm 
\epsfbox{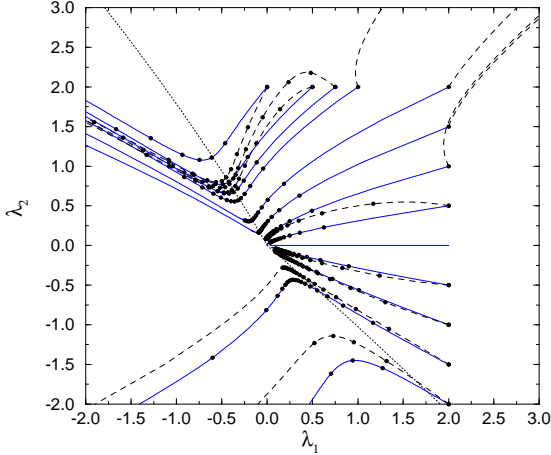}
\vspace{-1.3cm}
\caption{Perturbative RG trajectories starting from bare couplings
($\lambda_1(\Lambda), \lambda_2(\Lambda)$) along the lines $\lambda_1=2$
or $\lambda_2=2$. The solid lines (dash lines) correspond to one-loop
(two-loop) trajectories while the stability line is indicated as a
dotted line.  
Indicatively, the dots along a trajectory represent the evolution of
couplings after running by a factor of $e$ down to the infrared.}
\vspace{-0.5cm}
 \label{traj}
\end{figure}

Fig.~(\ref{traj}) displays the RG flows, at one-loop (solid lines) and
two-loop (dash lines) level, within renormalized perturbation theory. 
Then, starting from bare couplings $(\lambda_1(\Lambda),
\lambda_2(\Lambda))$,  if $\lambda_2 \neq 0$ all RG flows in the
infrared intersect the ``stability line'' \cite{YAM}, eventually
becoming runaway trajectories. The phase transition is then of first
order \cite{AMIT}. 
For $\lambda_2$ small, though, the flow is rather slow and even though
the ``stability line'' is crossed, this happens after many decades
of running; the phase transition is, in this case, weakly first order,
with $v \ll \Lambda$.
Along the $\lambda_2=0$ axis the transition is known to be second order,
and the well known triviality analysis \cite{TRIVIALITY} applies. The
two-loop corrections seem to improve the hierarchy $\Lambda/v$, as found
also in \cite{BHJ}. In particular, there exists a region with 
$\lambda_1,\lambda_2>0$  where the flow is towards larger values and it
appears that it never crosses the stability line.  However, this only
hints upon the breakdown of perturbation theory and a nonperturbative
analysis is called for.  
It is also essential that no other fixed point, unreachable in
perturbation theory, exists. Should one be present, the RG trajectories 
would be distorted and there could be regions where the transition is
second order. We found no evidence of such a fixed point. 

On the lattice, 
the physical parameter controlling the running of
the couplings and hence the size of the corrections is the correlation
length $\xi$ of the system. Then one expects that $\Lambda/v \sim
\xi^{\beta}$, where $\beta$ is the appropriate critical exponent: a
large hierarchy will only be possible if $\xi$ is big, or equivalently,
that the transition is weakly first order or second order.

\section{Monte Carlo Results}

\begin{table}[tb]
 \setlength{\tabcolsep}{1.pc}
\catcode`?=\active \def?{\kern\digitwidth}
\caption{Estimates of the jump in the order parameter $v^2$  evaluated
 at the critical point $m_c$, and of the  correlation length estimated by
 $\xi\simeq L^*/2$,  where $L^*$ is the  smallest lattice where
 coexistence was found, or  from the effective  potential.}  
\label{table}
\begin{center}
\begin{tabular}{cccc}
 \hline \hline
$(\lambda_1,\lambda_2)$ &  $m_c^2$ & $v^2$ & $\xi$ \\ \hline
(0.5, -0.45) & -0.772 & 0.83  &  7 \\
(-0.22, 0.5) & -0.91  & 2.10  &  3  \\ 
(0, 0.5)     & -2.42  &  0.5  & 40   \\ 
(-3.97, 8)   & -1.30  &  10-20      & $<$ 2   \\
(-14.97, 30) & -1.50  &  20-40     & $<$ 2   \\ 
(0, 8)       & -24.75   &  0.15     & 6   \\ 
(0, 16)      & -43.98 &  0.15  & 6 \\ 
(0, 30)      & -77.0  &  0.19  & 6    \\ 
(8, 8)       & -63.8  &  0.11   & 6    \\ 
(8, 16)      & -82.5  &  0.16   & 6   \\ 
(8, 30)      & -114.5 & 0.17   & 6  \\  \hline\hline
\end{tabular}
 \end{center}
\vspace{-0.8cm}
\end{table}

Table I shows all the points in the (bare) coupling constant space at
which we performed Monte Carlo simulations. We used two different
programs checked against each other: one based on a simple one-hit
Metropolis algorithm with a uniform step tuned so that the acceptance
rate is about 60$\%$ and the other based on the hybrid algorithm (with
or without Fourier acceleration). In this latter method, two parameters
have to be chosen, namely the number of leap-frog steps and the step size. 
This allows better control of the autocorrelation time. 
We found optimal CPU performance for 5-8 leap-frog steps before each
Metropolis test.  The hybrid one performed clearly better. 

We used as an order parameter (following \cite{YUE}) the expectation
value of the $U(N)\times U(N)$ invariant operator
$O=\mbox{\rm Tr}\, {\bar{\phi}}^\dagger {\bar{\phi}}$
which corresponds to the susceptibility, where 
$\bar{\phi}$ is the lattice average of each field component. $<O>$ 
is proportional to $v^2$ in the broken phase and zero in the
unbroken phase, modulo finite size corrections. 

We used lattices of sizes ranging from $L^4=4^4$ to $L^4=14^4$. 
In order to obtain information about the order of the transition, at
each given ($\lambda_1, \lambda_2$) we searched for hysteresis
effects in the measurement of the order parameter by performing thermal 
cycles in the relevant parameter, $m^2$, across the critical region.
Strong hysteresis loops is an indication of a strong first order
transition. 

On smaller lattices, ($4^4, 6^4, 8^4$),  the critical region was
identified by searching for a double-peak signal in the histogram
distribution of  ${\rm Tr}(\phi  \phi^{\dagger})$.
We then moved to bigger lattices ($10^4, 12^4, 14^4$) to look for
coexistence. 
Along the process of increasing the lattice  size we eventually begin to
see metastability at some size $L^*$. We estimate then the correlation
length to be $\xi \sim L^*/2$. Crude as this procedure may seem, it is
physically meaningful and it agrees, where comparison is possible, with
the effective potential.

\begin{figure}[tb]
 \vspace{-.2cm}
 \epsfxsize=80mm 
\epsfbox{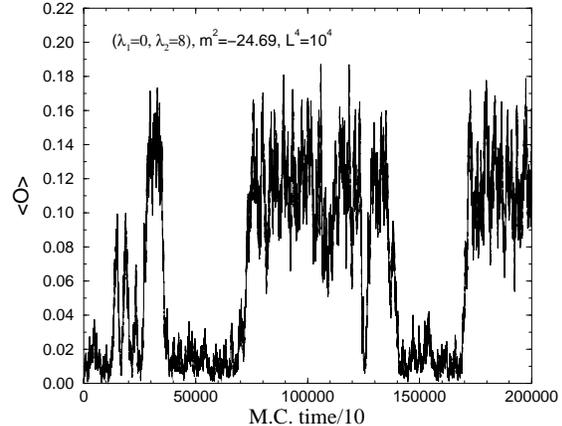}
 \vspace{-1.5cm}
\caption{Order parameter time history (over $2\times 10^6$ hybrid
sweeps) for the point (0, 8) on a $10^4$ lattice at $m^2=-24.69$.}
\label{fig08L10}
\end{figure}

Our results are as  follows. 
All points close to the stability line exhibited marked
hysteresis loops and hence show strong first order transitions,
becoming stronger as we move up along the stability line. For such
couplings, $\xi\simeq 1$ so the cut-off effects are big and the
connection to  continuum physics questionable. 
In the weak coupling region ($\lambda_1,\lambda_2 < 1$), we also
computed the one-loop bare effective potential \cite{YUE} and found that
it agreed with the numerical data within $10-30\%$.
Next we investigated couplings along the $\lambda_2$ axis. Typically,
runs on $4^4,6^4$ lattices did not show any hysteresis effects. 
However we found clear sign of the existence of two minima on  $10^4-14^4$
lattices.  
Fig.~(\ref{fig08L10}) displays tunneling between two phases for
the point (0, 8) on a $10^4$ lattice. For the same point, clear signal
was found of two coexistent minima on the $12^4$ lattice, but no
tunneling was observed. The transition becomes stronger with increasing
$\lambda_2$.  
For points deep in the $\lambda_1>0,\lambda_2>0$ region, we were able to
observe coexistence of phases, but only on $12^4, 14^4$ lattices. The
transition is always clearly first order, 
but characterized by correlation lengths, as expected, larger than those
obtained close to the stability line.
For couplings close to the $\lambda_1$ axis, our numerical results
are in agreement with the expectation of the RG for a weak first order
transition. In this case, though, it is difficult to distinguish a weak
first order from a second order transition. 

Our results are summarized in Table~(\ref{table}). From these,
one can see that in most of parameter space the v.e.v., $v(\Lambda)$,
is typically only one order of magnitude smaller than the cut-off. 
Our results are consistent with the standard perturbative picture of
first order phase transitions and the absence of any nontrivial fixed
point. The hierarchy $\Lambda/v$ is not ``tunable'' by $m^2$ as in the
$O(N)$ model, but rather depends on $\lambda_2$. 
Phenomenologically viable models must lead to couplings with small
$\lambda_2$ in order to support a large hierarchy.

\end{document}